\documentclass[12pt,a4paper]{article}
\usepackage{graphicx}
\usepackage{times}
\def\spose#1{\hbox to 0pt{#1\hss}}
\def\ltsimm{\mathrel{\spose{\lower 3pt\hbox{$\sim$}}
        \raise 2.0pt\hbox{$<$}}}
\def\gtsimm{\mathrel{\spose{\lower 3pt\hbox{$\sim$}}
        \raise 2.0pt\hbox{$>$}}}
\def\Mdot{\hbox{${\dot M}$}}

\def\km{{\rm\thinspace km}}
\def\cm{{\rm\thinspace cm}}

\def\s{{\rm\thinspace s}}
\def\yr{{\rm\thinspace yr}}
\def\g{{\rm\thinspace g}}
\def\kmps{\hbox{${\rm\km\s^{-1}\,}$}}

\def\erg{{\rm\thinspace erg}}

\def\Hz{{\rm\thinspace Hz}}

\def\ster{{\rm\thinspace ster}}
\def\ergps{\hbox{${\rm\erg\s^{-1}\,}$}}

\def\Msol{\hbox{$M_{\odot}$}}

\def\Msolpyr{\hbox{${\rm\Msol\yr^{-1}\,}$}}
\def\pcm{\hbox{${\rm\cm^{-1}\,}$}}
\def\pcm2{\hbox{${\rm\cm^{-2}\,}$}}
\def\pcm3{\hbox{${\rm\cm^{-3}\,}$}}
\def\ergpscm3Hz{\hbox{${\rm\ergps\cm^{-3}\Hz^{-1}\,}$}}
\def\ergpscm3Hzster{\hbox{${\rm\ergps\cm^{-3}\Hz^{-1}\ster^{-1}\,}$}}
\def\gpcm3{\hbox{${\rm\g\cm^{-3}\,}$}}
\def\ergpcm2{\hbox{${\rm\erg\cm^{-2}\,}$}}
\def\ergpcm3{\hbox{${\rm\erg\cm^{-3}\,}$}}
\def\phpscm2{\hbox{${\rm photons\s^{-1}\cm^{-2}\,}$}}

\title{\bf Theory and Models of Colliding Stellar Winds}
\author{J. M. Pittard\\
\vspace{1cm}\\
\normalsize School of Physics and Astrophysics, University of Leeds, United Kingdom
\\
\\
\normalsize Published in proceedings of \\
\normalsize"Stellar Winds in Interaction", editors T. Eversberg and J.H. Knapen. \\ 
\normalsize Full proceedings volume is available on http://www.stsci.de/pdf/arrabida.pdf
}
\date{\mbox{}}
\begin{document}
\maketitle
%
%
\def\bull{\vrule height .9ex width .8ex depth -.1ex}
\makeatletter
\def\ps@plain{\let\@mkboth\gobbletwo
\def\@oddhead{}\def\@oddfoot{\hfil\tiny\bull\quad
Workshop ``Stellar Winds in Interaction", Convento da Arr\'abida, 2010 May 29 - June 2 \quad\bull}%
\def\@evenhead{}\let\@evenfoot\@oddfoot}
\makeatother
%
%
\def\beginrefer{\section*{References}%
\begin{quotation}\mbox{}\par}
\def\refer#1\par{{\setlength{\parindent}{-\leftmargin}\indent#1\par}}
\def\endrefer{\end{quotation}}
%
%
{\noindent\small{\bf Abstract:} 
I discuss some of the important aspects of the phenomena of colliding
winds in massive binary star systems with a particular focus on WR\,140.
} 
%
%
\section{Introduction}
Luminous stars are able to drive powerful winds which, while ploughing
into their surroundings, may also cause significant mass loss and thus
modify the star's evolution. Determining the rate of mass loss is key
to understanding how such stars evolve. Fortunately, in massive binary
systems we are able to use the wind of one star as an in-situ probe of 
the wind of the other star. In addition, the collision of the winds unleashes
a broad spectrum of emission revealing the
interesting physics of high Mach number shocks, including the
acceleration of a small proportion of particles to
relativistic energies.

\section{The Dynamics of Colliding Wind Binaries}
\subsection{Instabilities}
\begin{figure}[ht]
\centering
\includegraphics[width=5.5cm]{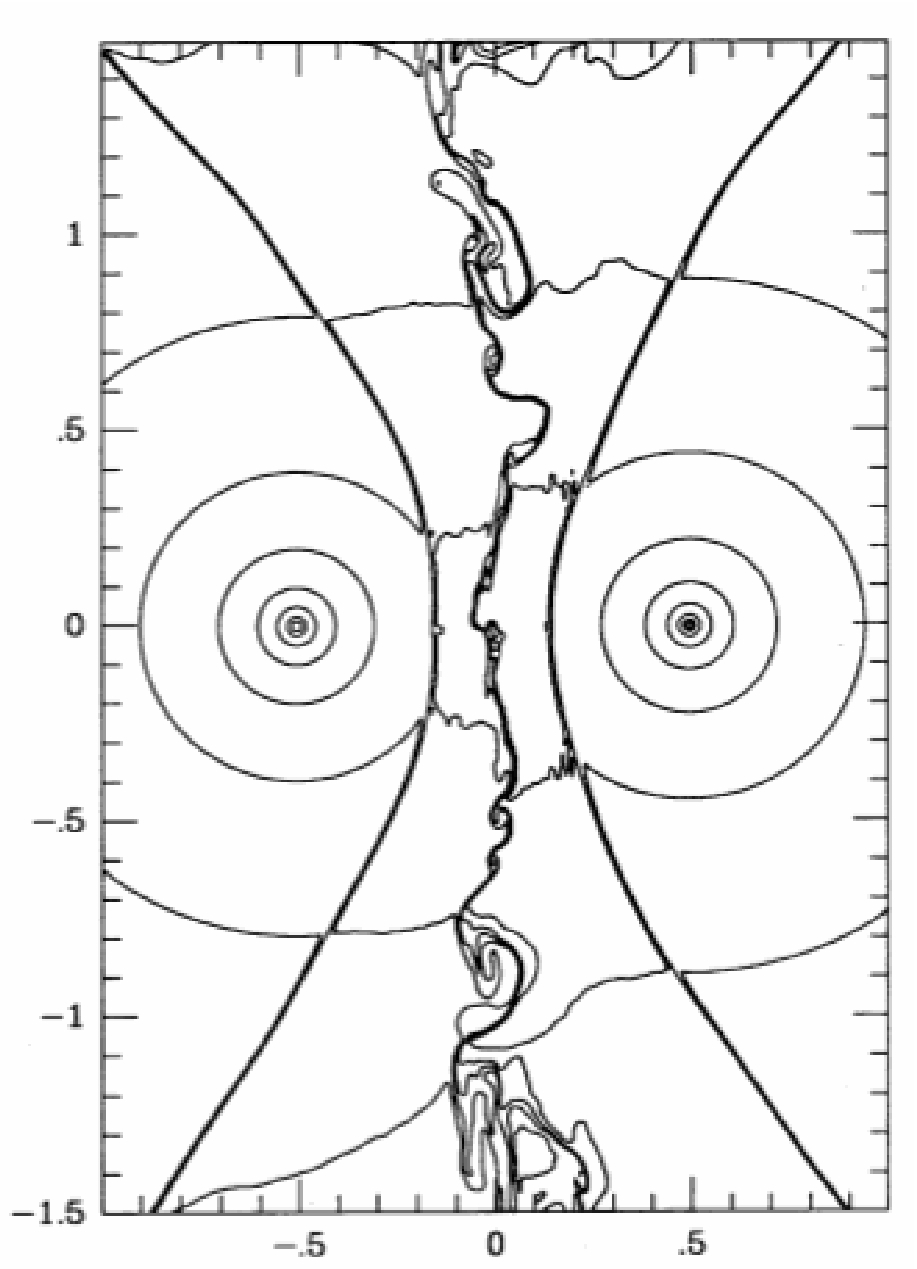}
\includegraphics[width=5.5cm]{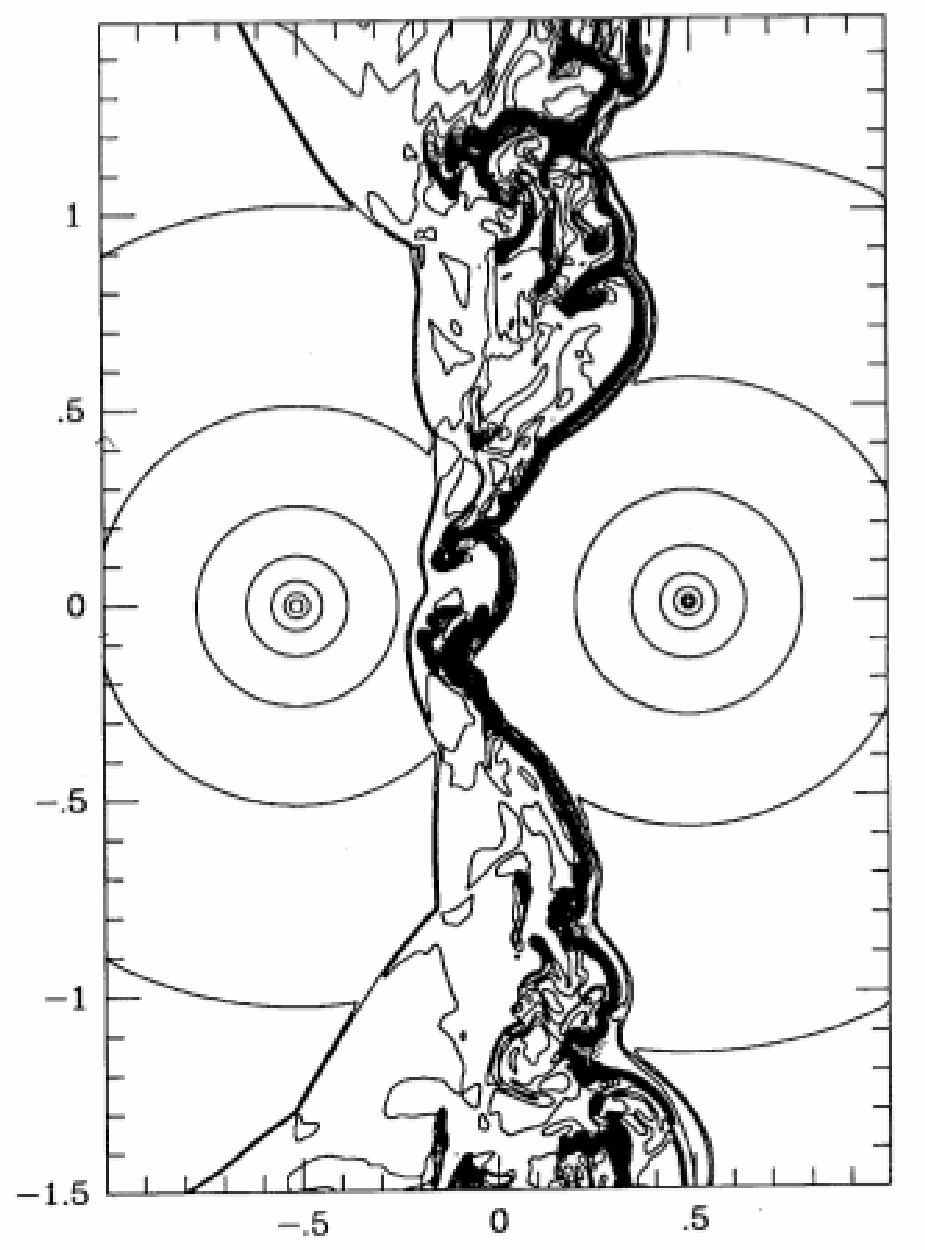}
\includegraphics[width=5.5cm]{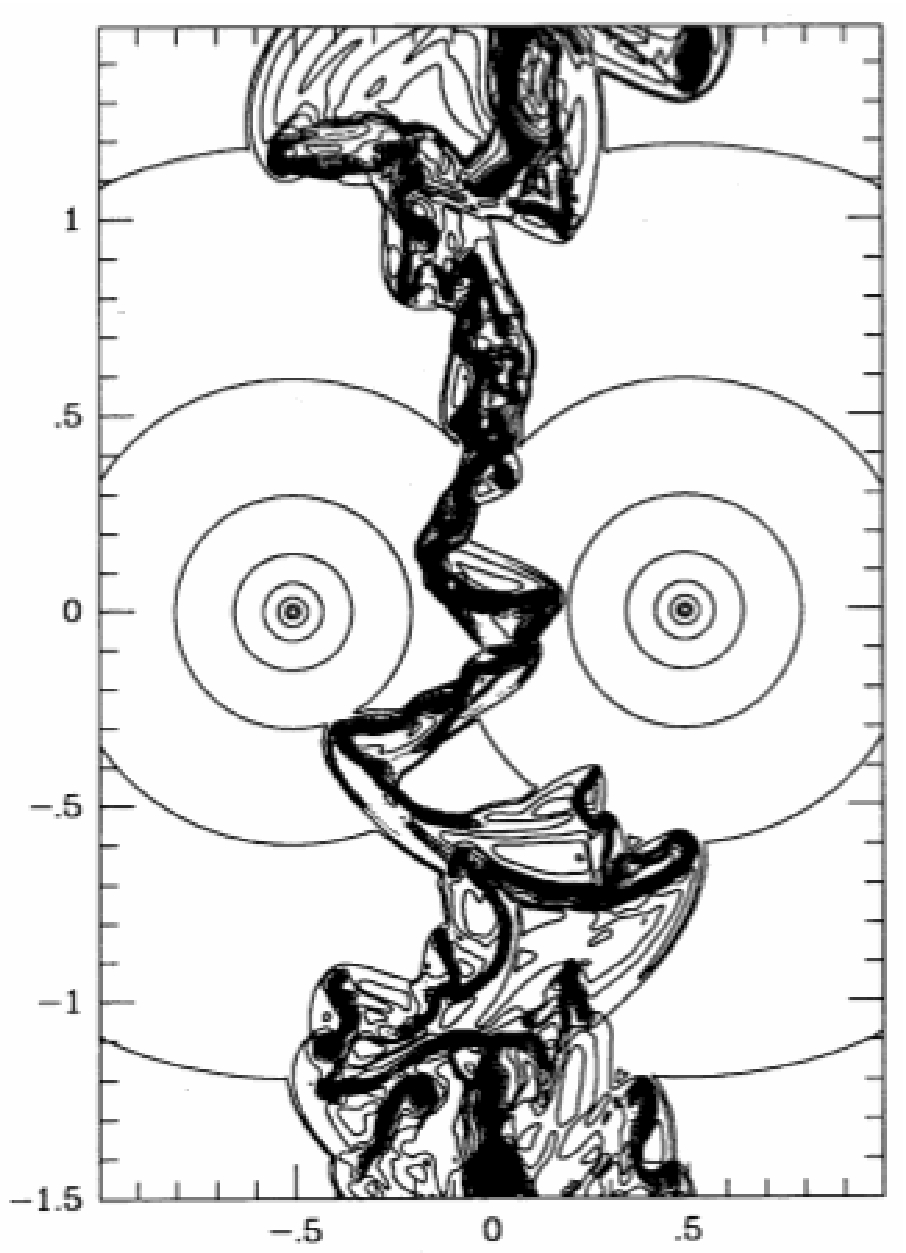}
\caption{Instabilities in the WCR of CWBs. {\it Left:} When both sides of
  the contact discontinuity are largely adiabatic, the WCR is very
  smooth.  {\it Center:} When one side is radiative, thin shell instabilities
  occur, but are somewhat limited by the ``cushioning'' of the hot gas
  (Vishniac 1983).  {\it Right:} When both sides are radiative, the much
  stronger and highly non-linear thin shell instability occurs
  (Vishniac 1994).  Adapted from Stevens et al. (1992).}
\label{fig:instabilities}
\end{figure}

The nature of the wind-wind collision in colliding wind binaries (CWBs) depends on a number of factors,
and displays a huge diversity. For instance, in some systems the winds
are of almost equal strength and collide roughly mid-way between the
stars.  In others, one wind is significantly more powerful than the
other and completely overwhelms the weaker wind, causing the collision
region to crash onto the surface of the companion star. Hence the
nature of the wind-wind collision region (WCR) depends on the wind properties of the system.
The thermal behaviour of the WCR also depends 
on the orbital properties of the system. In short period systems,
where the two stars are very close together, the wind-wind collision
is likely to be highly radiative. The shock-heated gas therefore cools
very rapidly, and a geometrically thin and dense region of gas forms
which is prone to severe, and perhaps disabling, non-linear
thin-shell instabilities (NTSI, Vishniac 1994)\footnote{This is 
the conventional wisdom, but in fact it is not clear
  exactly what occurs between the stars---e.g. even whether two ``winds''
  are produced---in such an extreme and hostile environment.}.  On the
other hand, if the orbital period is long, the shocked gas may behave
largely adiabatically, flowing out of the system while still hot. In
this case the WCR stays thick and
``puffed-up'', and is far less affected by disruptive instabilities, although
Kevin-Helmholtz instabilities, due to a velocity shear at the contact
discontinuity between the winds, may still occur. Where one
wind is radiative and the other largely adiabatic, a thin dense layer
of cooled gas abuts a thicker, hotter, but more rarefied layer which
acts like a ``cushion'' to damp out thin shell instabilities occurring
in the dense layer (Vishniac 1983). These differences were illustrated
by Stevens et al. (1992) and are reproduced in Fig.~\ref{fig:instabilities}.

The transition between radiative and adiabatic post-shock regions is
conveniently estimated using the value of $\chi \equiv t_{\rm
  cool}/t_{\rm dyn} \approx v_{8}^{4} D_{12}/\Mdot_{-7}$, where
$t_{\rm cool}$ is the cooling time of the gas, $t_{\rm dyn}$ is a
dynamical flowtime which is rather loosely defined but can be taken as
either the time for shocked gas at the apex of the WCR to flow a
distance $D_{\rm sep}$ downstream, or for the shocked gas from the
weaker wind to flow a distance $r_{\rm OB}$ downstream (see Stevens et
al. 1992 for these definitions). In the above, $v_{8}$ is the wind
speed normalised to $1000\,\kmps$, $D_{12}$ is the stellar separation
normalised to $10^{12}\,$cm, and $\Mdot_{-7}$ is the mass-loss rate
normalised to $10^{-7}\,\Msolpyr$. Note that this formalism for
$\chi$ depends on specific assumptions about the post-shock
temperature and the morphology of the cooling curve at this point (see
Pittard \& Stevens 2002). By specifying appropriate values for
$\Mdot_{-7}$, etc., it is possible to determine a separate value of
$\chi$ for each shocked wind.

One of the reasons why WR\,140 is of particular interest is that its
highly eccentric orbit results in big, but reproducible, changes in
the properties of the WCR, for instance its density.  While the shocked O-star wind is largely adiabatic throughout the
entire orbit, the same is not true of the shocked WR wind, which though 
adiabatic at apastron ($\chi_{\rm WR} \sim 50$), has much more significant cooling at periastron ($\chi_{\rm WR}
\sim 2$). At this time the shocked gas radiates away all of its energy
as it leaves the system, seemingly creating the ideal conditions for
dust formation (P.~ Williams, these proceedings).
A signature of this cooling is believed to be seen in the 
He\,{\sc i} 1.083-$\mu$m line, which develops a sub-peak
on top of its normal flat-topped profile as the stars approach periastron
(see Varricatt, Williams \& Ashok 2004, and also P.~ Williams, these proceedings).

The exact thermal behaviour of the shocked gas is impossible to
determine without fully three-dimensional numerical simulations, since
the value of $\chi$ as estimated above is only approximate. A
number of other mechanisms can also influence the value of $\chi$. For
instance, if the winds are clumpy, which is believed to be the case
for hot stars, the cooling and hence $\chi$ will be
underestimated. Furthermore, the efficient acceleration of non-thermal
particles in the WCR may sap energy from the thermal
plasma, again causing it to cool more rapidly. Therefore, values of
$\chi$ estimated using the above equation should be taken as only a
rough guide of the true thermal behaviour.

\subsection{Radiative Driving Effects}

Systems where the stars are relatively close to each other (i.e. those
with short periods and/or highly eccentric orbits) may suffer from
interactions between one star's wind and the other star's radiation
field.  Two scenarios have been determined. In the ``radiative
inhibition'' scenario, the acceleration of one or both winds is
inhibited by the radiation field of the other star, thus reducing the
speed of the wind(s) before the collision (Stevens \& Pollock
1994). Another effect, termed ``radiative braking'' comes into play in
systems where the stronger wind closely approaches the more luminous
star (Gayley et al. 1997). This condition is met in many Wolf-Rayet
(WR) + O-star binaries. The stronger radiation field may then
efficiently couple to the more powerful wind, causing a sudden
deceleration or braking. The effect is highly non-linear, and the
braking can be so severe that the stronger wind can be prevented from
colliding with the surface of the more luminous star even in cases
where a normal ram-pressure balance between the winds would not be
possible. Neither of these effects is believed to be important in
WR\,140, since even at periastron the stars are still well separated
(see R.~ Fahed et al., these proceedings),
and the WCR is at least several stellar radii from the surface of
the O-star despite the large disparity in wind strengths.

\subsection{Orbital Effects}

The orbit of the stars can also strongly influence the properties and
nature of the WCR. The most obvious effect is 
a spiralling of the WCR as the gas within it flows out
of the system. Due to the large computational cost of 3D simulations,
it is only recently that the first 3D hydrodynamical models of
colliding wind binaries were presented in a refereed journal (Lemaster
et al. 2007), and this work lacked several key
processes which are important in the short-period systems where the
effects of orbital motion are greatest.  The most notable omissions
were the lack of any treatment for the acceleration of the winds and
the cooling of the shocked gas.  Shortly afterwards, 3D
smoothed-particle hydrodynamics (SPH) simulations of the WCR in the
massive binary system $\eta$~Carinae were presented (Okazaki et
al. 2008). Though these models were isothermal, and did not solve for
the temperature structure behind the shock, they provided much insight
into the dynamics of the WCR in this highly eccentric ($e \approx
0.9$) system.  At about the same time, a ``dynamic'' model 
was presented by Parkin \& Pittard (2008). This model did not solve the
hydrodynamic equations, but instead mapped the apex of the WCR (given by
the equations in Stevens et al. 1992) into a 3D space. The apex was provided
with a time-dependent skew which aimed to reflect the ratio of the
wind to orbital speeds, and the gas was assumed to behave ballistically
further downstream. Though the resulting dynamics are only representative 
of the true situation, a comparison against results from a full hydrodynamical
calculation revealed that this approach does a more than adequate job
in many situations.

\begin{figure}[ht]
\centering
\includegraphics[width=16.0cm]{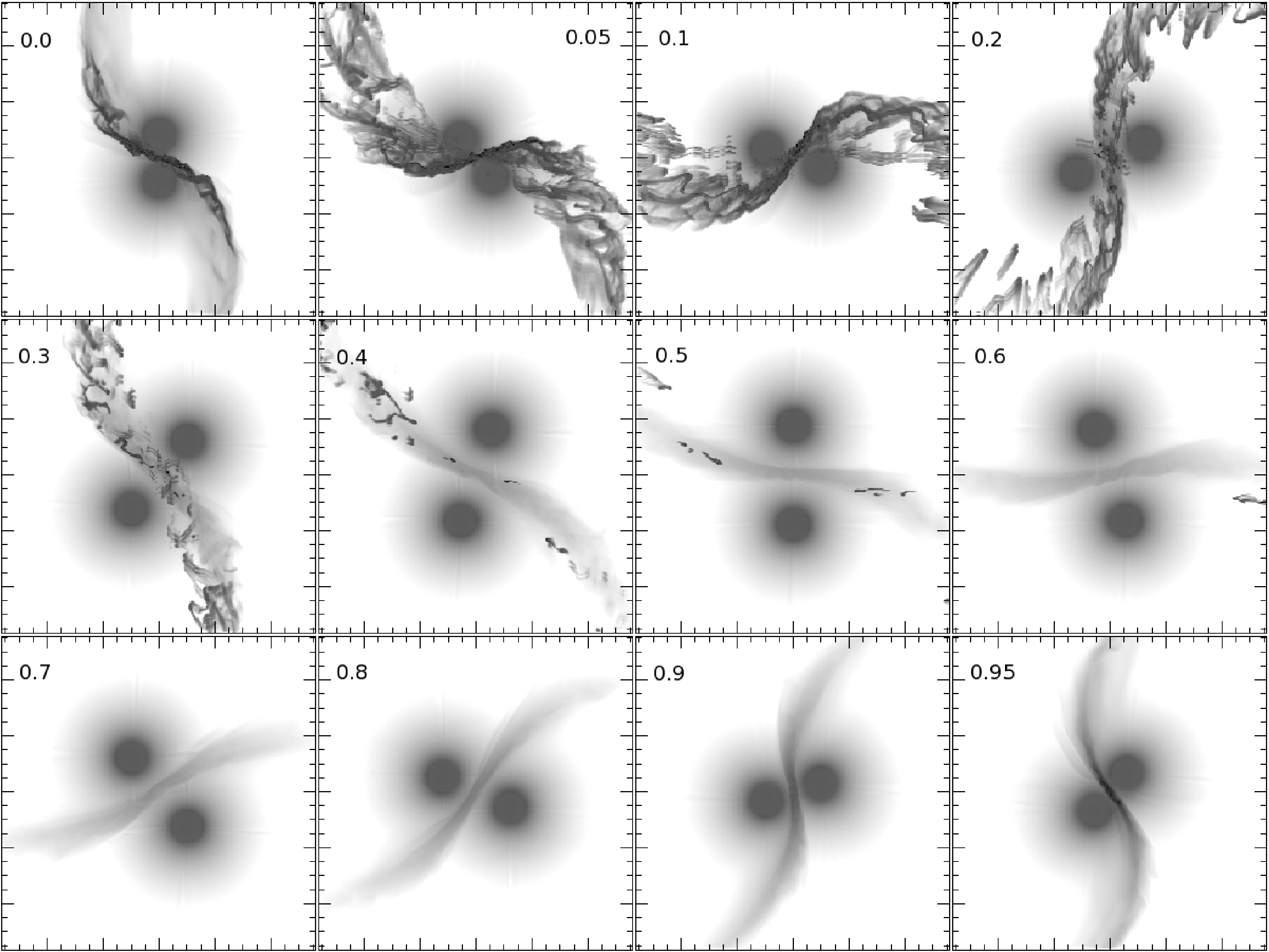}
\caption{Intensity images at 1000\,GHz for an observer viewing
an eccentric ($e=0.36$) O+O-star system with an orbital
period of 6.1\,d (adapted from Pittard 2010). 
See Pittard (2009) for details of the hydrodynamical model).}
\label{fig:hystersis}
\end{figure}

The first 3D simulations of CWBs to include orbital motion, the
radiative driving of the winds and cooling of the shocked gas were
presented by Pittard (2009).  Focusing on O+O-star systems, models
with circular orbits were presented. Depending on the orbital period
(3\,d or 10\,d) the WCR was either highly radiative or largely
adiabatic. In the former, differences in the nature of the
instabilities in the leading versus the trailing edge of each ``arm''
of the WCR are seen. In the latter, the density and temperature of the
hot gas changes across the arms. The most interesting behaviour,
however, was seen from a system with an eccentric orbit where the
stellar separation at periastron was the same as
the system with the 3\,d orbital period, while the apastron
separation was the same as the system with the 10\,d orbital period -
this required $e=0.36$.  Most notably, the properties of the WCR exhibited a
distinct phase lag compared to those expected from the instantaneous
position of the stars, which reflects the recent history of the stellar
separation. Hence there were marked differences in the WCR properties
when the stars are at identical stellar separations depending on
whether the stars are approaching or receding from each other. For
instance, the gas in the WCR remains hot until near phase 0.9, after
which it collapses into a thin dense sheet which is torn apart by
instabilities. Yet it is past apastron before the cold clumps are
cleared away from the stars (this long timescale is due to the high
inertia of the dense clumps relative to the rarefied gas which flows
past them). This process is clearly shown in synthetic images at
1000\,GHz (see Fig.~\ref{fig:hystersis}).

For most of WR\,140's orbit, orbital effects are minimal, and the WCR
is approximately axisymmetric, with rotational symmetry about the line
running through the centres of the stars. However,
orbital effects become very significant as the stars swoop through
periastron passage. The rapid changes in the positions of the
stars induces severe curvature into the WCR, breaking the axisymmetry
which exists up until about phase 0.95. One must then be careful
when interpreting data using axisymmetric models for the WCR (e.g. the
L\"{u}hrs 1997 model), since this curvature will bias estimates
of the opening angle of the WCR, and thus of the momentum ratio of the winds.

\subsection{The Effect of Clumpy Winds}
Studies of how wind clumping affects the WCR have been presented by
Walder (1998), who showed that dense clumps can tip an otherwise
marginally adiabatic WCR into a radiative regime.  Pittard (2007)
examined the effect of clumps on the adiabatic WCR of WR\,140 at
apastron. If the clumps are not too dense or large (so that they do
not punch through the WCR), they can be rapidly destroyed by the
vorticity created during their passage through the shocks bounding the
WCR. Although the WCR becomes highly turbulent as a result, the
overall effect is to smooth out the flow. Thus determinations of the
stellar mass-loss rates using the X-ray emission from a WCR may be
relatively insensitive to clumping, and thus offer a useful
alternative to other methods where this is not the case. The strong
turbulence occurring within the WCR also has implications for particle
acceleration and the mixing of the winds.

\section{Models of WR\,140}
Due to its interesting behaviour, WR\,140 is one of the best studied
CWB systems, and this has led to the development of a large number of
theoretical models. Each model has typically addressed one aspect of
its emission. In the following I summarise the models which have been
made of the thermal X-ray and non-thermal radio emission. The latter
have also been used to predict the expected flux of the high-energy
non-thermal emission, so I will discuss these too. I will not discuss
the thermal IR emission from WR\,140, which has been extensively
studied and modelled by Peredur Williams and collaborators, details of
which are given in the contribution by Peredur Williams in these
proceedings.

\subsection{Thermal X-ray Emission} 
WR\,140 is an exceptionally luminous X-ray source for a colliding wind
binary system, with an X-ray luminosity $\sim 4\times10^{34}$\,erg\,s$^{-1}$.  
This indicates that, in addition to the strong,
dense and fast wind from the Wolf-Rayet star, the companion star must
also have a powerful wind (the maximum efficiency for converting wind
power into hot shocked gas is obtained when the winds are of equal
strength).  The X-ray luminosity displays large, phase-repeatable,
variations around the orbit. Such variations must reflect the underlying changes
occurring to the hot plasma in the WCR, augmented by changes in the
circumstellar absorption from the surrounding winds (see, e.g., M.~Corcoran et
al., these proceedings).

To model the emission from the WCR with hydrodynamic codes, one must
resolve the cooling length behind the shock. This is a difficult task
when the cooling is very rapid, but perfectly possible for WR\,140
since the shocked gas is never strongly radiative. Early predictions
for the X-ray spectrum and lightcurve were made by Stevens et
al. (1992).  The predicted X-ray luminosity scales as $D^{-1}$, where
$D$ is the stellar separation, and there is a deep X-ray minimum as
the orbit moves the denser wind of the WR star in front of the apex of
the WCR.  Prior to this work, a simple point-source model for the
X-ray absorption was constructed to compare against the observed
absorption of EXOSAT spectra (Williams et al. 1990).  Today,
observations by the {\it RXTE} X-ray satellite reveal the X-ray
lightcurve in exquisite detail (see M.~Corcoran et al., these
proceedings). An immediate puzzle is that the expected $D^{-1}$ scaling
is not observed---instead the real increase is lower.  This has yet to
be explained, but may be related to the cooling of the plasma or the
acceleration of non-thermal particles within the WCR. Furthermore,
models which adopt spatially extended X-ray emission at the WCR are not able
to reproduce the deep X-ray minimum (see, e.g., M.~Corcoran et al., these
proceedings).

Several other models have investigated various physical aspects of the
high Mach number shocks found in WR\,140. The heating of electrons and
ions at the shocks was modelled by Zhekov \& Skinner (2000) who
concluded that models where the electrons were heated more slowly than
the ions provided better fits to the {\it ASCA} X-ray data available
at the time (see also Pollock et al. 2005). Another process which is
likely to have a significant timescale is that of ionisation
equilibrium---the initial post-shock plasma is under-ionised for its
temperature, and ionisation towards equilibrium proceeds as material
flows downstream from the shock and eventually out of the system. This
effect is beautifully illustrated in Pollock et al. (2005), where it
is seen that the FWHM of various X-ray lines increases with the
ionisation potential of the species. This is the opposite trend to
that expected from a plasma in collisional ionisation equilibrium
(CIE), where the hottest gas and most highly ionised species occur at
the apex of the WCR where the shocks are perpendicular to the oncoming
winds.  Pollock et al. (2005) estimates that the distance for
ionisation equilibrium to be established when the stars are at apastron
is about $32$\,AU (i.e. comparable to the orbital separation). Models
of X-ray line profiles (see Henley, Stevens \& Pittard 2003; Henley et
al. 2005, 2008) where the plasma is assumed to be in CIE fail
spectacularly when an attempt is made to fit the WR\,140 data. This
adds further support to the plasma being in non-equilibrium
ionisation.

By constructing a number of hydrodynamical models and examining the
expected X-ray emission from each, Pittard \& Dougherty (2006)
determined a range of possible mass-loss rates and wind momentum ratios
which were consistent with the observed X-ray flux. The mass-loss
rates should be accurate to $\pm 25$\%, with uncertainties due to the
ill-determined composition of the WR wind (see Pollock et al. 2005 and
Pittard \& Dougherty 2006 for more details). Zhekov \& Skinner (2000)
derived slightly higher mass-loss rates for their adopted wind
momentum ratio ($\eta = \Mdot_{\rm O}v_{\rm O}/\Mdot_{\rm WR}v_{\rm
  WR} = 0.0353$) as they assumed relatively low abundances for C and
O.  Note that previous works have sometimes used mass-loss rates which
are inconsistent with the X-ray luminosity (e.g. Dougherty et
al. 2005). The values of $\eta$ investigated by Pittard \& Dougherty
(2006) spanned the range $0.02-0.2$. Larger values of $\eta$ require
relatively smaller values of $\Mdot_{\rm WR}$ and larger values of
$\Mdot_{\rm O}$. The latest investigation on the sudden increase in
the absorption component of the He\,{\sc I} 1.083-$\mu$m line near
periastron implies that $\eta=0.1$ (see the contribution by Peredur
Williams in these proceedings), which in turn implies that $\Mdot_{\rm
  WR} \approx 2\times10^{-5}\,\Msolpyr$ and $\Mdot_{\rm O} \approx
2\times10^{-6}\,\Msolpyr$. Further (3D) modelling is needed to test
these values.

\subsection{Non-Thermal Radio Emission}

In the late 1970s and early 1980s the radio emission from WR\,140 was
determined to be highly variable, with both the flux and the spectral
index undergoing significant changes. Over the intervening years
further observations in the radio have revealed in exquisite detail
phase repeatable light curves and images of the emission from the WCR
(see Williams et al. 1990; White \& Becker 1995; Dougherty et
al. 2005; S.~Dougherty et al., these proceedings). Part of the
modulation is thought to be caused by the variable circumstellar
extinction to the source of the non-thermal (synchrotron) emission
(the WCR) as the O-star orbits in and out of the radio photosphere in
the dense WR wind. However, the intrinsic non-thermal emission
probably also varies around the orbit. Various aspects of the particle
acceleration in and non-thermal emission from CWBs were discussed by
Eichler \& Usov (1993).

Early models of the non-thermal radio emission from CWBs were very
simple. It was usually assumed that the observed flux ($S_{\nu}^{\rm
  obs}$) was a combination of the free-free flux from the spherically
symmetric winds ($S_{\nu}^{\rm ff}$), plus the flux from a point-like
non-thermal source located at the stagnation point of the WCR
($S_{\nu}^{\rm nt}$). In this model the non-thermal emission is then
attenuated by free-free absorption (opacity $\tau_{\nu}^{\rm ff}$)
through the surrounding winds:
\begin{equation}
\label{eq:simplerad}
S_{\nu}^{\rm obs} = S_{\nu}^{\rm ff} + S_{\nu}^{\rm nt}\,{\rm e}^{-\tau_{\nu}^{\rm ff}}.
\end{equation}
While this approach allows simple solutions to the radiative transfer
equation (e.g., Williams et al. 1990; Chapman et al. 1999), such models
fail to reproduce the spectral variation of the emission with orbital
phase. Williams et al. (1990) therefore proposed that in future models
of WR\,140 the low-opacity ``hole'' in the dense WR wind created by
the O-star's wind should be accounted for. However, White \& Becker
(1995) pointed out that in WR\,140, even the O-star's wind has
significant opacity. Together, these works demonstrated the need for
more realistic models which account for the spatial extent of the
emission and absorption from the circumstellar envelope and the
WCR. More realistic models should also account for the effects of
various cooling mechanisms (e.g. inverse Compton, adiabatic, etc.) on
the non-thermal electron distribution, and also additional absorption
mechanisms (e.g., the Razin effect).

The modelling of the non-thermal radio emission from CWBs took a
dramatic jump forward when key assumptions in previous models, such as
a point-like source of non-thermal emission, and a spherically
symmetric, single temperature, surrounding envelope, were removed.
This was achieved in Dougherty et al. (2003), where models of the
thermal and non-thermal radio emission used 2D, cylindrically
symmetric, hydrodynamical simulations, to more accurately describe the
density and temperature structure of the system. Sight-lines to the
observer could now pass through regions of both high and low
opacity. The assumption of a point-source of non-thermal emission was
also removed, by treating the emission in a phenomenological way
where the local density of non-thermal electrons and the magnetic
field strength were related to the local thermal energy density within
the WCR. The non-thermal electrons were further assumed to have a
power-law distribution, $N(\gamma)d\gamma = C \gamma^{-p}d\gamma$,
where $\gamma$ is the Lorentz factor, $C$ is a proportionality
constant which fixes the normalisation, and it was assumed that $p=2$
(suitable for test particle diffusive shock acceleration, with strong shocks and an adiabatic
index equal to $5/3$).

Although this work was not directly applied to WR\,140, the results provided
a great deal of new insight into the nature of the radio emission from
CWBs. An immediate realisation was the potential importance of the
Razin effect in attenuating the low frequency synchrotron emission
within the WCR. Several key scaling relationships were also
established. For instance, in the absence of IC cooling the total
synchrotron emission from the entire WCR in adiabatic systems was
found to scale as $D^{-1/2}\nu^{-1/2}$ (as a reminder the X-ray
emission in the optically thin, adiabatic limit, scales as
$D^{-1}$). In a successive paper, Pittard et al. (2006) investigated
the effect of inverse Compton (IC) cooling of the non-thermal
electrons as they flowed downstream from their accelerating shocks and
out of the system. This work showed that with IC cooling the {\em
  intrinsic} luminosity actually {\em declines} with stellar
separation. The effect of the stellar separation on the {\em thermal}
radio flux was also explored. It was discovered that the {\em thermal}
radio emission from the WCR scales as $D^{-1}$, in an identical way to
the thermal X-ray emission. Since this emission is optically thin in
systems with an adiabatic WCR, it can mimic a synchrotron component,
so that one should rather cautiously interpret data with a spectral
index $-0.1 \ltsimm \alpha \ltsimm 0.5$ (Pittard et al. 2006).

\begin{figure}[ht]
\centering
\includegraphics[width=8.0cm]{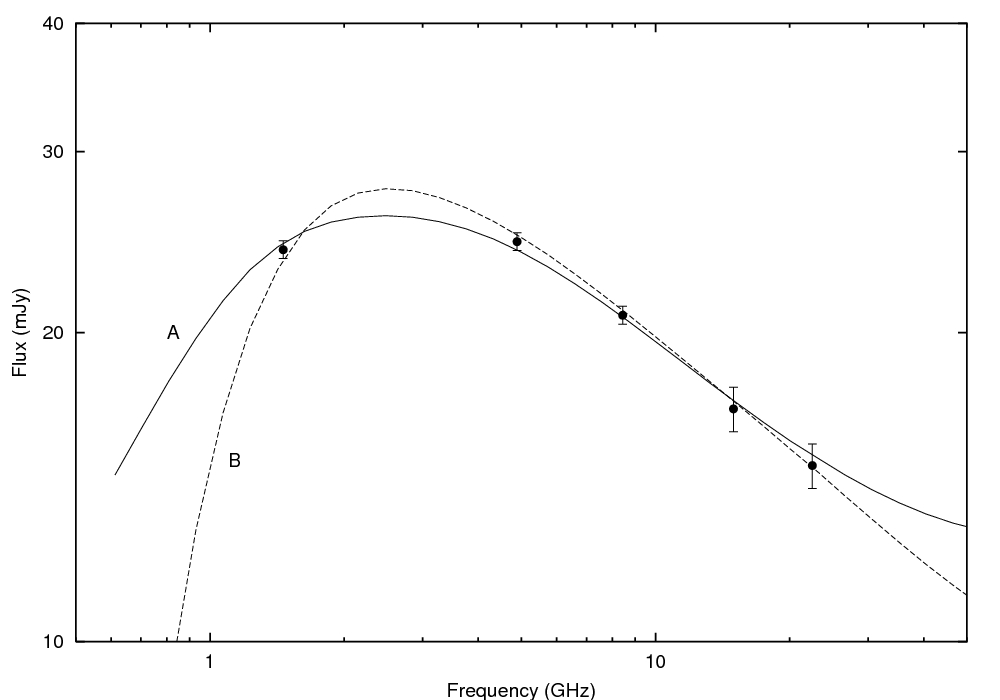}
\includegraphics[width=8.0cm]{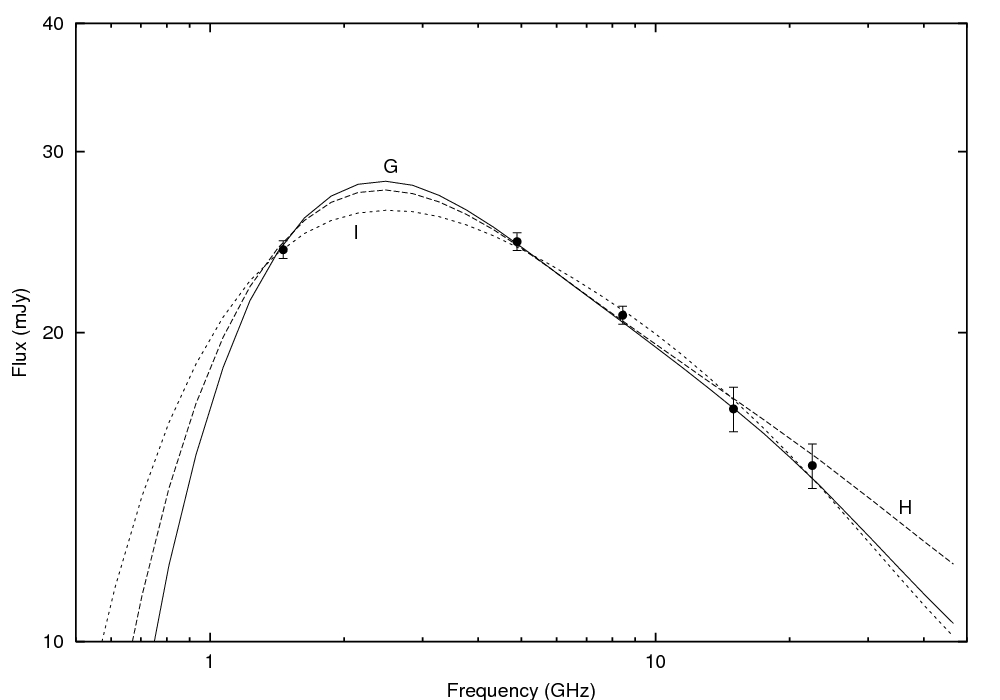}
\caption{Model fits to the radio data of WR\,140 at $\phi=0.837$.
{\it Left}: Fits where the low-frequency turndown is due to free-free absorption.
Models A and B span plausible values of the wind momentum ratio, $\eta$.
{\it Right}: Fits where the Razin effect is responsible for the turndown.
For further details of the models see Pittard \& Dougherty (2006).}
\label{fig:wr140fits}
\end{figure}

The model of Pittard et al. (2006) was applied to WR\,140 in Pittard
\& Dougherty (2006). Fits were obtained to data at orbital phase
0.837, which is around the peak of the non-thermal radio
lightcurve. It was found that the low frequency turndown in the
radio spectrum could be explained as either free-free absorption
through the surrounding stellar winds, or the Razin effect (see
Fig.~\ref{fig:wr140fits}). In the former case it proved impossible to
obtain a good fit to the data with $p=2$.  The best fit was obtained
with $p=1.4$, though changes to the assumed magnetic field strength
allow some small variation in this value. Such indices can result from the
shock re-acceleration process, whereby the non-thermal particles pass
through a sequence of shocks (Pope \& Melrose 1994), or from 2nd order
Fermi acceleration. Either of these processes may be significant in
CWBs, since the clumpy nature of the winds means that the WCR is
likely to be highly turbulent, with weak shocks distributed throughout
it (Pittard 2007). In contrast, fits with the Razin effect dominant do
allow $p=2$, though require a worryingly high efficiency of electron
acceleration. For this reason, fits with free-free absorption dominant
were preferred. A wide range of wind momenta could fit the data in
this case, though it might be possible to constrain the models with
future, high sensitivity VLBA observations. 

\subsection{Non-Thermal X-ray and $\gamma$-ray Emission}

The presence of non-thermal electrons and ions should produce 
non-thermal emission at X-ray and $\gamma-$ray energies via several
mechanisms, including the up-scattering of lower energy (e.g., stellar UV)
photons (inverse Compton [IC] scattering), relativistic bremsstrahlung, and pion decay.
Early predictions for the high-energy non-thermal emission from CWBs 
were made by Benaglia \& Romero (2003). In this work it was assumed that
IC scattering was the dominant mechanism. Then the
ratio of the luminosity from IC scattering to the synchrotron luminosity
is equal to the ratio of the energy density of the target photons, $U_{\rm ph}$,
to the magnetic field energy density, $U_{\rm B}$:
\begin{equation}
\frac{L_{\rm ic}}{L_{\rm sync}} = \frac{U_{\rm ph}}{U_{\rm B}}.
\end{equation}
While very straightforward, the predictive power of this equation is
severely curtailed by the fact that the magnetic field strength in the WCR is
generally very uncertain. Since $U_{\rm B} \propto B^{2}$, small changes in
the estimated value of $B$ lead to large changes in $L_{\rm ic}$.

\begin{figure}[ht]
\centering
\includegraphics[width=5.5cm]{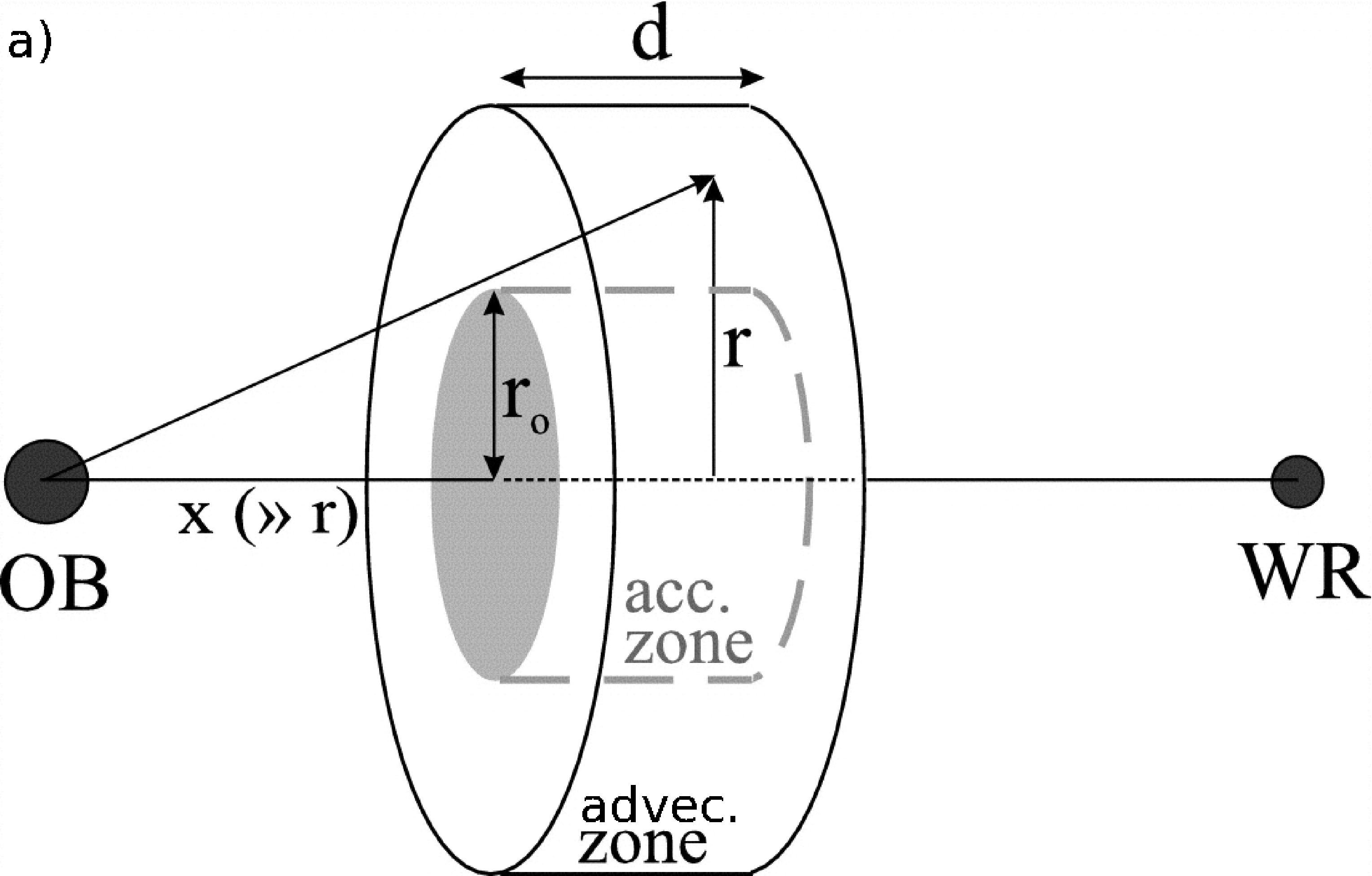}
\includegraphics[width=5.5cm]{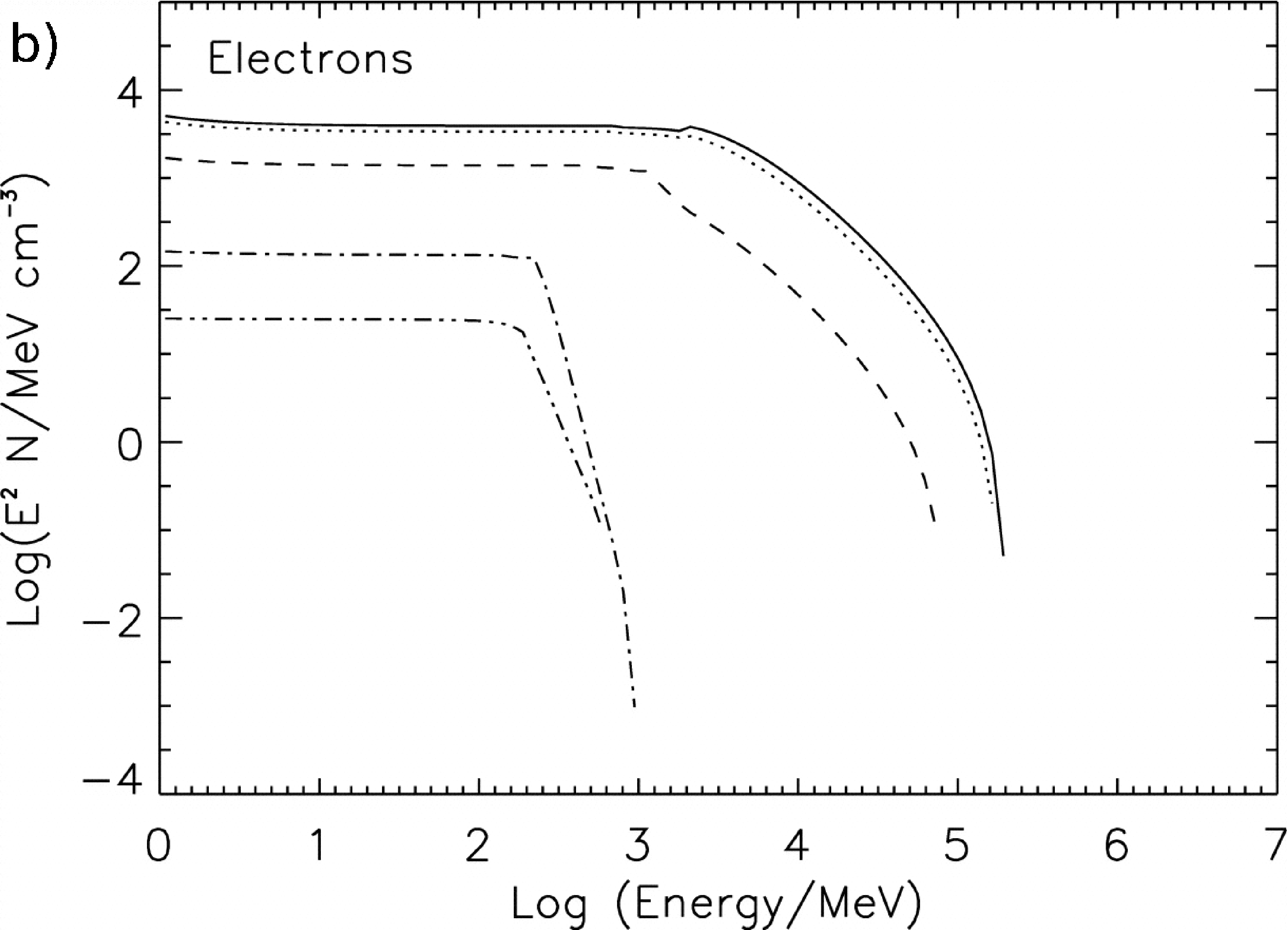}
\includegraphics[width=5.5cm]{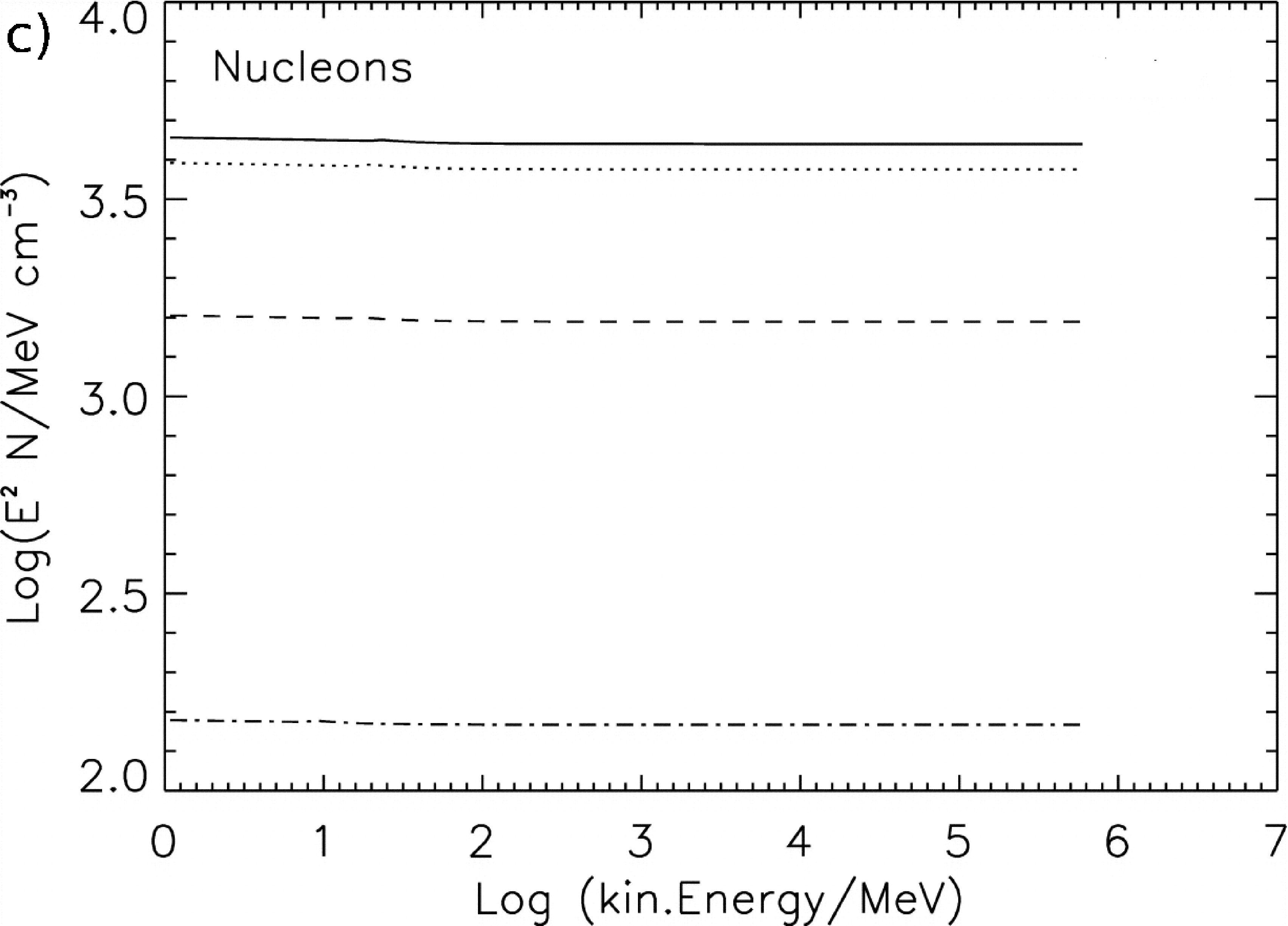}
\caption[]{a) Geometry of the 2-zone model in Reimer et al. (2006). b) Evolution
of the non-thermal electron spectrum from the inner acceleration zone ({\it solid
line}) as a function of downstream distance in the advection zone. At low 
energies adiabatic/expansion losses dominate, while at high energies IC
losses dominate. c) As b) but for nucleons. Only adiabatic losses occur.
See Reimer et al. (2006) for further details.}
\label{fig:2zone}
\end{figure}

\begin{figure}
\centering
\includegraphics[width=8.0cm]{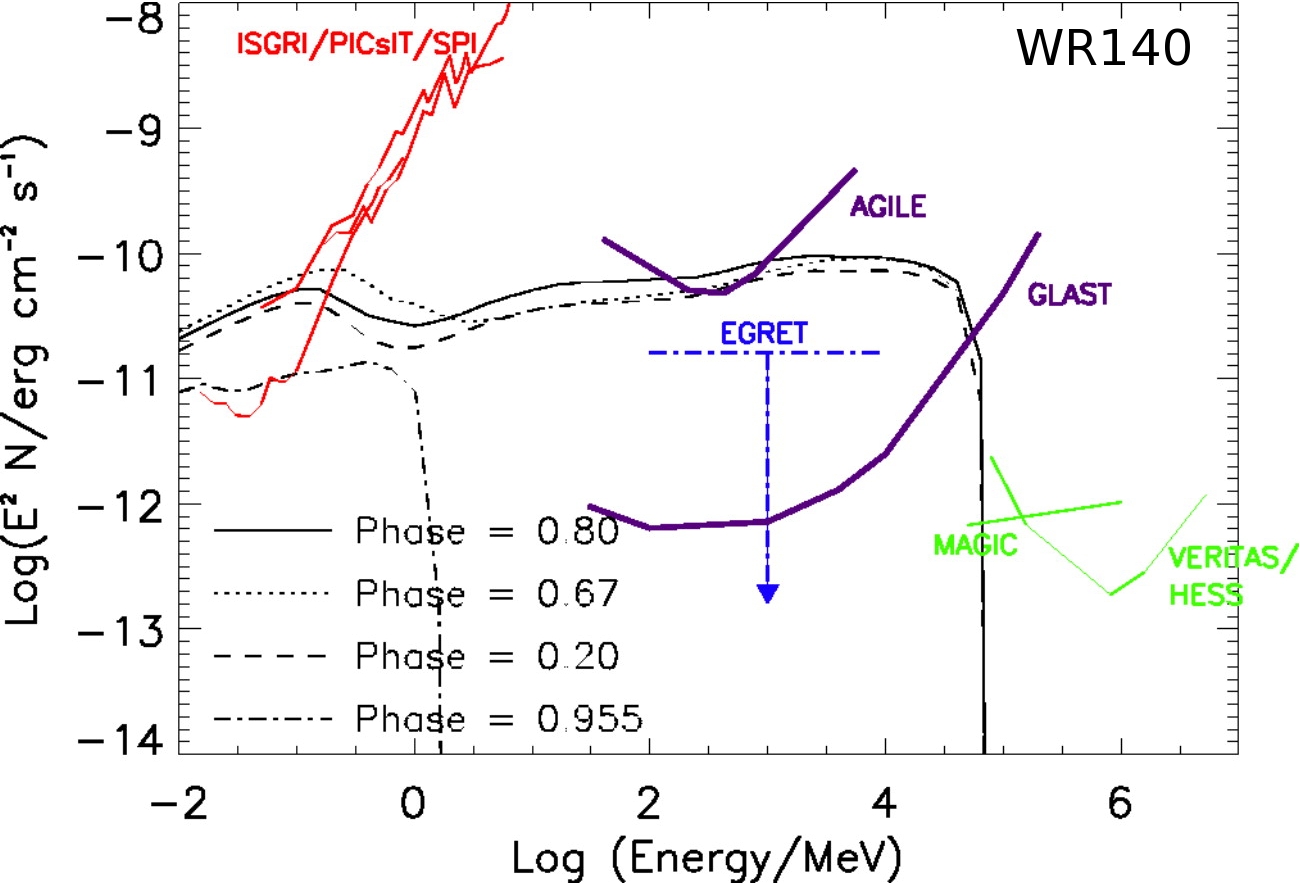}
\includegraphics[width=8.0cm]{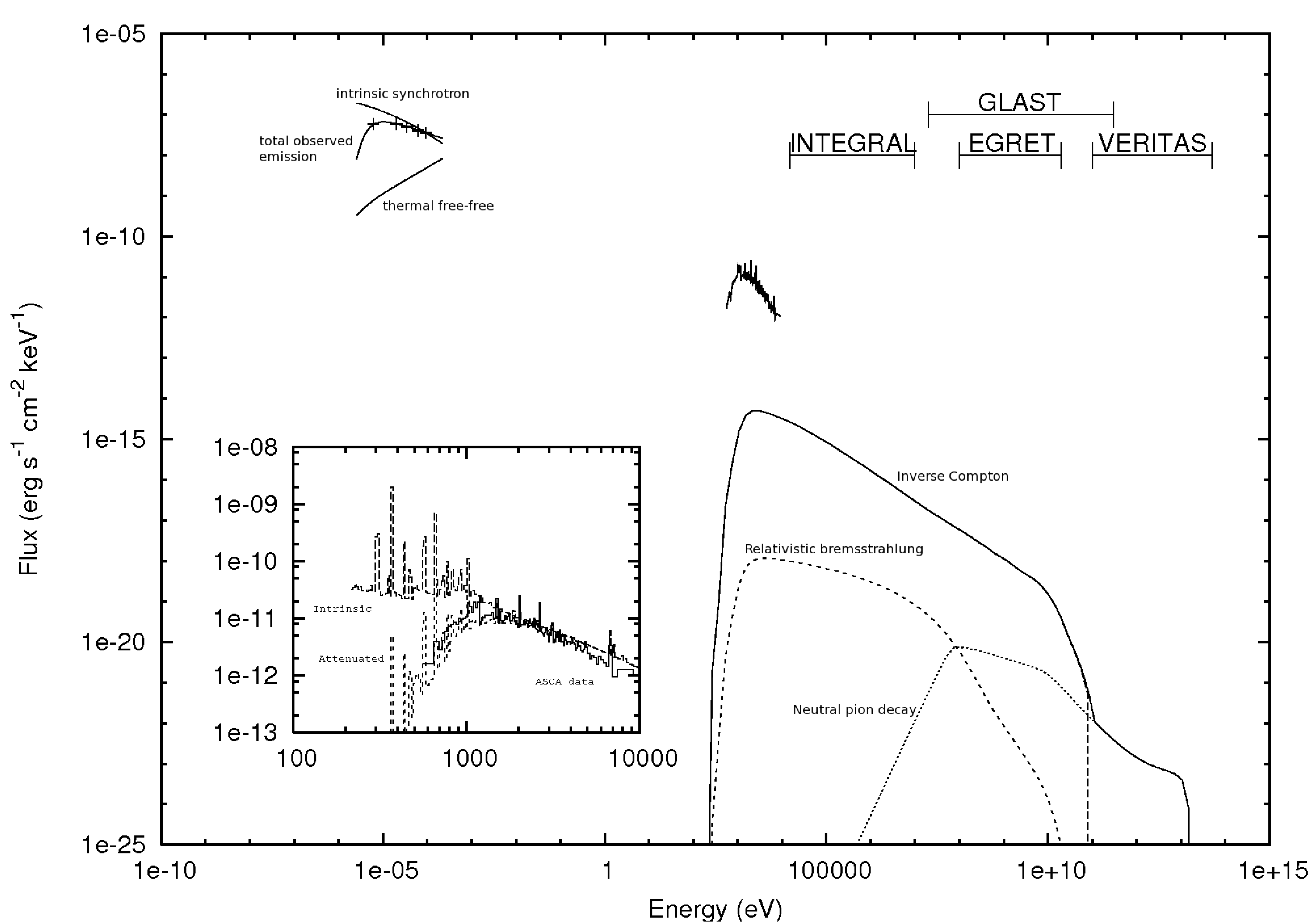}
\caption[]{{\it Left:} Predicted IC spectra for WR\,140 at phases 0.2,
  0.67, 0.8 and 0.955 from Reimer et al. (2006). $\gamma$-ray
  absorption is not included.  {\it Right:} The radio and non-thermal
  UV, X-ray and $\gamma$-ray emission calculated from model~B in
  Pittard \& Dougherty (2006), together with the observed radio and
  X-ray flux (both at orbital phase 0.837). The model IC ({\it long dash}),
  relativistic bremsstrahlung ({\it short dash}), and neutral pion decay
  ({\it dotted}) emission components are shown, along with the total
  emission ({\it solid}).  See Pittard \& Dougherty (2006) for more
  details.}
\label{fig:wr140he}
\end{figure}

In recent years the dramatic sensitivity gains achieved by space-based
satellites and ground-based arrays of Cerenkov telescopes have raised
the tantalising prospect of the first detection of the non-thermal
X-ray and $\gamma$-ray emission from CWBs (in fact, we now believe
that we have detected the massive CWB $\eta$\,Carinae at $\gamma$-ray
energies, Abdo et al. 2010; Walter et al. 2010). This, in turn, has led to new
theoretical predictions (e.g., Bednarek 2005). The anisotropic nature
of the IC process, where the emitted power is dependent on the
scattering angle, was considered by Reimer, Pohl \& Reimer
(2006). This work developed a two-zone model of the non-thermal
emission. Particles are accelerated in an inner zone where their
spatial diffusion exceeds their motion due to advection with the
background fluid. Their energy distribution is self-consistently
computed by considering all relevant gain and loss mechanisms.
Particles are assumed to be resident within this region until their
timescales for advection and diffusion are comparable, after which
they are assumed to move into the advection region where they suffer
further losses as they flow downstream. Fig.~\ref{fig:2zone} shows the
assumed geometry and the resultant non-thermal energy spectra of the
electrons and nucleons.  The left panel of Fig.~\ref{fig:wr140he}
shows the predicted IC emission from WR\,140 as a function of orbital
phase. While Reimer et al. (2006) conclude that while WR\,140 should
be easily detected with GLAST/Fermi, the high galactic background has
meant that unfortunately no detection has yet been made. The change in
the IC flux with viewing angle due to anisotropic scattering is likely
to be obscured by large variations in the energy density of the
stellar radiation fields resulting from the high orbital eccentricity.

Pittard \& Dougherty (2006) also predicted the non-thermal X-ray and
$\gamma$-ray emission from WR\,140, using fits to the radio and
thermal X-ray emission as constraints. The adopted approach was quite
different but complementary to that of Reimer et al. (2006)---while
the non-thermal particle spectrum was assumed rather than calculated,
and the IC scattering was treated as isotropic, it benefitted from a
more realistic description of the density and temperature distribution
within the system based on the X-ray constraints. Pittard \& Dougherty
showed that the uncertain particle acceleration efficiency and
spectral index have at least as much influence on the predicted flux
as the angle-dependence of the IC emission.

The right panel of Fig.~\ref{fig:wr140he} shows a predicted spectral
energy distribution for WR\,140 from one of the models presented in
Pittard \& Dougherty (2006).  Large differences in the predicted
$\gamma$-ray emission occur depending on whether the low frequency
turndown in the radio spectrum results from free-free absorption
through the surrounding stellar winds, or from the Razin effect.
While Pittard \& Dougherty (2006) could not determine the nature of
the absorption process from the fits to the radio spectrum, future
$\gamma$-ray detections will determine the $\gamma$-ray flux and
spectral index, and thus will also distinguish the nature of the
low-frequency turndown. The acceleration efficiency of the non-thermal
electrons and the strength of the magnetic field will then both be
revealed.

\section{Conclusions}
Theoretical models of the hydrodynamics of and emission from CWBs,
including of WR\,140, continue to improve. Hydrodynamical models are
becoming more sophisticated, with the first 3D simulations to be
published in refereed journals being made in only the last several
years. The goal now is to gradually add further important physics,
such as non-equilibrium ionisation and electron heating, to these 3D
models. Predictions for the non-thermal emission have also progressed,
leaving behind some of the assumptions in earlier works. Further
progress will be made when the non-thermal particle distribution
function is self-consistently solved with an accurate description of
the post-shock flow and cooling.

%
%
\section*{Acknowledgements}
I would like to express my gratitude to the workshop organisers
and to the amazing observational programme performed by the
amateur astronomers. This work is supported by a University
Research Fellowship from the Royal Society, UK.

%
%
\footnotesize
\beginrefer

\refer Abdo A.~A., et al., 2010, arXiv:1008.3235
 
\refer Bednarek W., 2005, MNRAS, 363, L46

\refer Benaglia P., Romero G.~E., 2003, A\&A, 399, 1121

\refer Chapman J.~M., Leitherer C., Koribalski B., Bouter R., Storey M., 1999, ApJ, 518, 890

\refer Dougherty S.~M., Beasley A.~J., Claussen M.~J., Zauderer B.~A., Bolingbroke N.~J., 2005, ApJ, 623, 447

\refer Dougherty S.~M., Pittard J.~M., Kasian L., Coker R.~F., Williams P.~M., Lloyd H.~M., 2003, A\&A, 409, 217

\refer Gayley K.~G., Owocki S.~P., Cranmer S.~R., 1997, ApJ, 475, 786

\refer Henley D.~B., Corcoran M.~F., Pittard J.~M., Stevens I.~R., Hamaguchi K., Gull T.~R., 2008, ApJ, 680, 705

\refer Henley D.~B., Stevens I.~R., Pittard J.~M., 2003, MNRAS, 346, 773

\refer Henley D.~B., Stevens I.~R., Pittard J.~M., 2005, MNRAS, 356, 1308

\refer Lemaster M.~N., Stone J.~M., Gardiner T.~A., 2007, ApJ, 662, 582

\refer L\"{u}hrs S., 1997, PASP, 109, 504

\refer Okazaki A.~T., Owocki S.~P., Russell C.~M.~P., Corcoran M.~F., 2008, MNRAS, 388, L39

\refer Parkin E.~R., Pittard J.~M., 2008, MNRAS, 388, 1047

\refer Pittard J.~M., 2007, ApJ, 660, L141

\refer Pittard J.~M., 2009, MNRAS, 396, 1743

\refer Pittard J.~M., 2010, MNRAS, 403, 1633

\refer Pittard J.~M., Dougherty S.~M., 2006, MNRAS, 372, 801

\refer Pittard J.~M., Dougherty S.~M., Coker R.~F., O'Connor E., Bolingbroke N.~J., 2006, A\&A, 446, 1001

\refer Pittard J.~M., Stevens I.~R., 2002, A\&A, 388, L20

\refer Pollock A.~M.~T., Corcoran M.~F., Stevens I.~R., Williams P.~M., 2005, ApJ, 629, 482

\refer Pope M.~H., Melrose D.~B., 1994, PASA, 11, 175

\refer Reimer A., Pohl M., Reimer O., 2006, ApJ, 644, 1118

\refer Stevens I.~R., Blondin J.~M., Pollock A.~M.~T., 1992, ApJ, 386, 265

\refer Stevens I.~R., Pollock A.~M.~T., 1994, MNRAS, 269, 226

\refer Varicatt W.~P., Williams P.~M., Ashok N., 2004, MNRAS, 351, 1307

\refer Vishniac E.~T., 1983, ApJ, 274, 152

\refer Vishniac E.~T., 1994, ApJ, 428, 186

\refer Walder R., 1998, Ap\&SS, 260, 243

\refer Walter R., Farnier C., Leyder J.-C., 2010, arXiv:1008.2533

\refer White R.~L., Becker R.~H., 1995, ApJ, 451, 352

\refer Williams P.~M., van der Hucht K.~A., Pollock A.~M.~T., Florkowski D.~R., van der Woerd H., Wamsteker W.~M., 1990, MNRAS, 243, 662

\refer Zhekov S.~A., Skinner S.~L., 2000, ApJ, 538, 808

\endrefer

\end{document}